%% file: main.tex
\documentclass[10pt, preprint2]{emulateapj}
\usepackage{verbatim}
\usepackage{amsmath}
\usepackage[usenames, dvipsnames]{color}

\newcommand{\planet}{WASP-18b}

\newcommand{\rsun}{\ensuremath{R_\sun}}
\newcommand{\msun}{\ensuremath{M_\sun}}

\newcommand{\mjup}{\ensuremath{M_{\rm Jup}}}
\newcommand{\teff}{\ensuremath{T_{\rm eff}}}

\newcommand{\kms}{\ensuremath{\rm km\,s^{-1}}}
\newcommand{\ms}{\ensuremath{\rm m\,s^{-1}}}
\newcommand{\rstar}{\ensuremath{R_s}}
\newcommand{\mstar}{\ensuremath{M_s}}

\newcommand{\mic}{\ensuremath{\mu \rm m}}

\newcommand{\tess}{{\it TESS}}

\newcommand{\sig}[1]{\ensuremath{#1\sigma}}
\newcommand{\figr}[1]{Figure~\ref{fig:#1}}
\newcommand{\secr}[1]{Section~\ref{sec:#1}}
\newcommand{\eqr}[1]{Eq.~\ref{eq:#1}}
\newcommand{\tabr}[1]{\mbox{Table~\ref{tab:#1}}}

\shorttitle{WASP-18 \textit{TESS} phase curve}
\shortauthors{Shporer et~al.}

\begin{document}

\title{\textit{TESS} full orbital phase curve of the WASP-18\MakeLowercase{b} system}

\author{
Avi Shporer\altaffilmark{1},
Ian Wong\altaffilmark{2,}\altaffilmark{$\dag$},
Chelsea X.~Huang\altaffilmark{1,}\altaffilmark{$\ddag$},
Michael R.~Line\altaffilmark{3}, 
Keivan G.~Stassun\altaffilmark{4,5},  
Tara Fetherolf\altaffilmark{6}, 
Stephen R.~Kane\altaffilmark{7}, 
Luke G.~Bouma\altaffilmark{8}, 
Tansu Daylan\altaffilmark{1,}\altaffilmark{*}, 
Maximilian N.~G{\"u}enther\altaffilmark{1,}\altaffilmark{$\ddag$}, 
George R.~Ricker\altaffilmark{1}, 
David W.~Latham\altaffilmark{9}, 
Roland Vanderspek\altaffilmark{1}, 
Sara Seager\altaffilmark{2}, 
Joshua N.~Winn\altaffilmark{8}, 
Jon M.~Jenkins\altaffilmark{10}, 
Ana Glidden\altaffilmark{1,2},  
Zach Berta-Thompson\altaffilmark{11},  
Eric B.~Ting\altaffilmark{10}, 
Jie Li\altaffilmark{10,12}, 
Kari Haworth\altaffilmark{1} 
}
\altaffiltext{1}{Department of Physics and Kavli Institute for Astrophysics and Space Research, Massachusetts Institute of Technology, Cambridge, MA 02139, USA}
\altaffiltext{2}{Department of Earth, Atmospheric, and Planetary Sciences, Massachusetts Institute of Technology, Cambridge, MA 02139, USA}
\altaffiltext{3}{School of Earth \& Space Exploration, Arizona State University, Tempe, AZ 85287, USA} 
\altaffiltext{4}{Vanderbilt University, Department of Physics \& Astronomy, 6301 Stevenson Center Lane, Nashville, TN 37235, USA}
\altaffiltext{5}{Fisk University, Department of Physics, 1000 17th Avenue N., Nashville, TN 37208, USA}
\altaffiltext{6}{Department of Physics and Astronomy, University of California, Riverside, CA 92521, USA}
\altaffiltext{7}{Department of Earth Sciences, University of California, Riverside, CA 92521, USA}
\altaffiltext{8}{Department of Astrophysical Sciences, Princeton University, Princeton, NJ 08544, USA}
\altaffiltext{9}{Harvard-Smithsonian Center for Astrophysics, 60 Garden Street, Cambridge, MA 02138, USA}
\altaffiltext{10}{NASA Ames Research Center, Moffett Field, CA 94035, USA},
\altaffiltext{11}{Department of Astrophysical and Planetary Sciences, University of Colorado, Boulder, CO 80309, USA}
\altaffiltext{12}{SETI Institute, Moffett Field, CA 94035, USA}
\altaffiltext{$\dag$}{\textit{51 Pegasi b} Fellow}
\altaffiltext{$\ddag$}{Juan Carlos Torres Fellow}
\altaffiltext{*}{Kavli Fellow}
\begin{abstract}

We present a visible-light full orbital phase curve of the transiting planet WASP-18b measured by the \tess\ Mission. The phase curve includes the transit, secondary eclipse, and sinusoidal modulations across the orbital phase shaped by the planet's atmospheric characteristics and the star-planet gravitational interaction. We measure the beaming (Doppler boosting) and tidal ellipsoidal distortion phase modulations and show that the amplitudes of both agree with theoretical expectations. We find that the light from the planet's day-side hemisphere occulted during secondary eclipse, with a relative brightness of $341_{-18}^{+17}$~ppm, is dominated by thermal emission, leading to an upper limit on the geometric albedo in the \tess\ band of 0.048 (\sig{2}). We also detect the phase modulation due to the planet's atmosphere longitudinal brightness distribution. We find that its maximum is well-aligned with the sub-stellar point, to within 2.9~deg (\sig{2}). We do not detect light from the planet's night-side hemisphere, with an upper limit of 43~ppm (\sig{2}), which is 13\% of the day-side brightness. The low albedo, lack of atmospheric phase shift, and inefficient heat distribution from the day to night hemispheres that we deduce from our analysis are consistent with theoretical expectations and similar findings for other strongly irradiated gas giant planets. This work demonstrates the potential of \tess\ data for studying full orbital phase curves of transiting systems. Finally, we complement our study by looking for transit timing variations (TTVs) in the \tess\ data and combined with previously published transit times, although we do not find a statistically significant TTV signal.

\end{abstract}

\section{Introduction}
\label{sec:intro}

The high-sensitivity data provided by visible-light space-based transit surveys, designed to detect the minute decrease in flux as a planet transits across its host star, can be used to look for variability along the entire orbital phase of a star-planet system. Beyond the transit, the phase curve includes the secondary eclipse, when the planet's day-side hemisphere is occulted by the host star, and sinusoidal brightness modulations across the orbital phase. 

While the transit depth is sensitive primarily to the planet-star radius ratio, the secondary eclipse depth is determined by the planet's thermal emission and the geometric albedo in the observed bandpass. Modulations along the orbital phase, in visible light, are shaped by the gravitational interaction between the star and planet, as well as the longitudinal variations of the planet's brightness. 

More specifically, the shape of the measured phase curve is a superposition of the effects of four main processes, described briefly and somewhat simplistically below, where we assume the transit to be at zero orbital phase, the orbit to be circular, and the planet's rotation to be synchronized with the orbit (i.e., tidally locked), as expected for short period systems \citep{mazeh2008}:
(1) Beaming, or Doppler boosting, where the periodic red- and blue-shifting of the host star's spectrum observed in the bandpass follows its orbital motion around the system's center of mass \citep[e.g.,][]{shakura1987, loeb2003, zucker2007, shporer2010}. The shape of the photometric variability reflects the orbital radial velocity (RV) curve, albeit with the opposite sign, and therefore can be described as a sine at the orbital period.
(2) Tidal distortion of the host star by the planet \citep[e.g.,][]{morris1985, morris1993, pfahl2008, jackson2012}, which leads to a cosine modulation at the first harmonic of the orbital period, commonly referred to as ellipsoidal modulation.
(3) Thermal emission from the planet's atmosphere, where a tidally-locked planet's day-side hemisphere (facing the star) is hotter than the planet's night-side hemisphere (facing away from the star), resulting in a cosine modulation at the orbital period. This process dominates the phase curve modulations in the near-infrared \citep[e.g.,][]{knutson2012, wong2015, wong2016}, but for highly irradiated planets it can also be detected at visible wavelengths \citep{snellen2009}.
(4) Stellar light reflected by the planet's atmosphere, which due to the geometric configuration of the system reaches maximum at the phase of secondary eclipse and minimum at mid-transit, producing a cosine variation at the orbital period \citep[e.g.,][]{jenkins2003, shporer2015}. Both the thermal emission and reflected light modulations (processes 3 and 4 above) are expected to have the same schematic shape, but a different amplitude. Hence, when combined they are commonly referred to as the atmospheric phase component \citep{parmentier2017}. 

The summary above shows that the orbital phase curve is sensitive to the star-planet mass ratio and characteristics of the planet's atmosphere, including the geometric albedo and thermal emission, along with their longitudinal distribution. For a review of visible-light orbital phase curves, see \citealt{shporer2017} and references therein. 

We present here our analysis of the \tess\ orbital phase curve of \planet\ (TIC  100100827, TOI 185, \citealt{hellier2009, southworth2009}). This massive 10.4~\mjup\ gas giant planet orbits its host star at a very short orbital period of 0.94 days, which establishes favorable conditions for a strong phase curve signal at visible wavelengths and allows us to study the atmosphere of a gas giant planet with high surface gravity subjected to high stellar irradiation. Our analysis of the visible-light orbital phase curve adds to previous measurements in the near-infrared of the phase curve \citep{maxted2013} and the secondary eclipse \citep{nymeyer2011, maxted2013, sheppard2017, arcangeli2018}.

The \tess\ observations are described in \secr{obs}, and our data analysis is described in \secr{dataanal}. We present our results in \secr{res} and discuss their implications in \secr{dis}. We conclude with a brief summary in \secr{sum}.

\section{Observations}
\label{sec:obs}

\begin{figure}
\includegraphics[width=\linewidth]{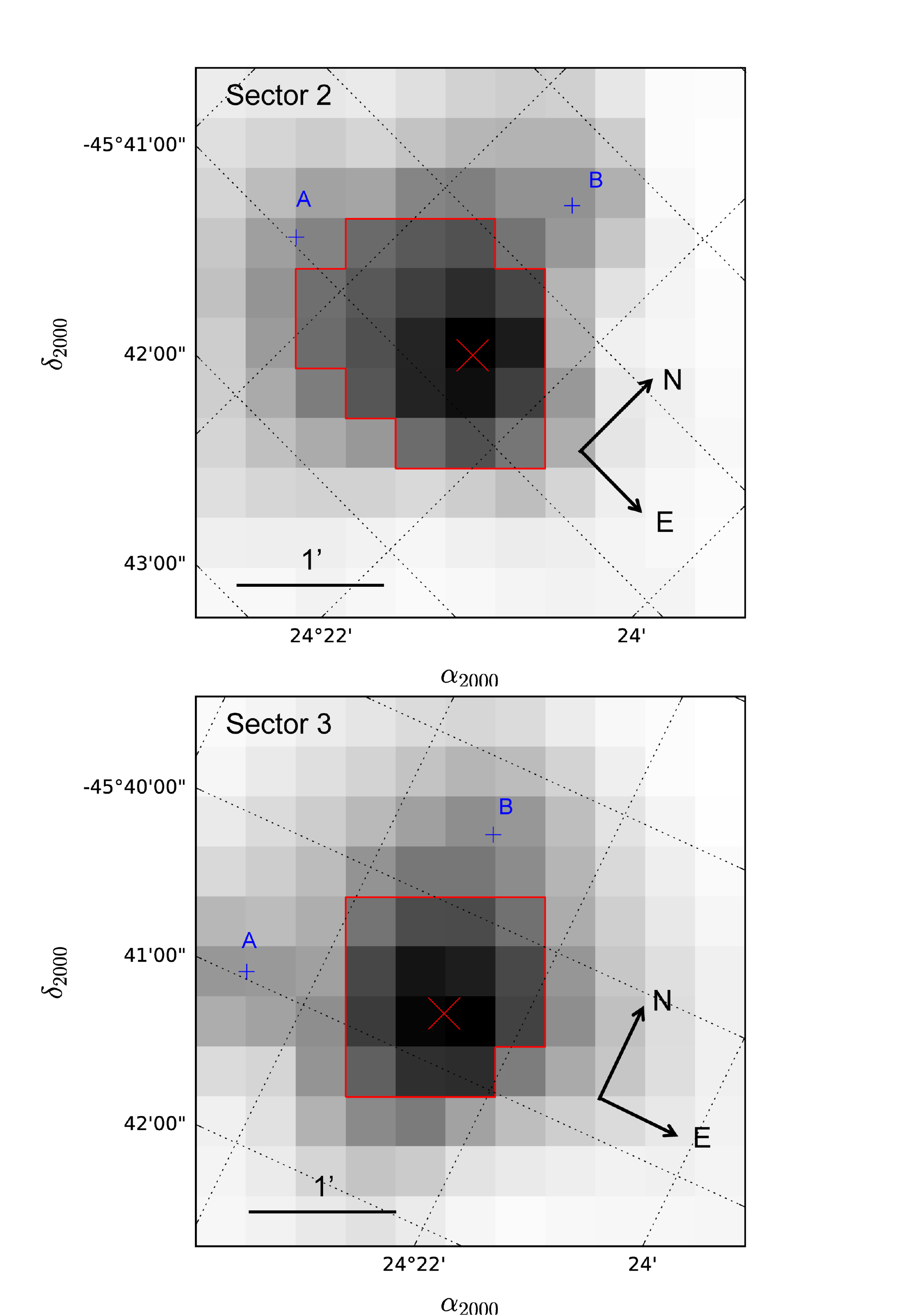}
\caption{Stacked \tess\ images of the 11$\times$11~pixel stamps centered around WASP-18 (red cross) for Sectors 2 and 3. The photometric extraction aperture used by the SPOC pipeline is outlined in red. The positions of two nearby $T\sim 12.5$ mag stars, which have been deblended by the SPOC pipeline, are denoted with blue crosses: TIC 100100823 (A) and TIC 100100829 (B).
}
\label{fig:stamp}
\end{figure}

The WASP-18 system was observed by Camera 2 of the \tess\ spacecraft during Sector 2 (from 2018 August 22 to 2018 September 20) and Sector 3 (from 2018 September 20 to 2018 October 18). WASP-18 is listed in the \tess\ input catalog (TIC; \citealt{stassun2018}) as ID 100100827 and included in the list of pre-selected target stars, which are observed with a 2-minute cadence using an 11$\times$11 pixel subarray centered on the target. The photometric data were processed through the Science Processing Operations Center (SPOC) pipeline \citep{jenkins2016}, hosted at the NASA Ames Research Center, which is largely based on the predecessor \textit{Kepler} mission pipeline \citep{jenkins2010, jenkins2017}. The stacked \tess\ subarrays produced by the SPOC pipeline are shown in Figure~\ref{fig:stamp}. Outlined in red are the optimal apertures used to extract the WASP-18 light curve in each sector.

For the results presented in this paper, we use the Presearch Data Conditioning (PDC, \citealt{smith2012, stumpe2014}) light curves from the SPOC pipeline. The data files include quality flags that indicate when photometric measurements may have been affected by non-nominal operating conditions on the spacecraft or may otherwise yield unreliable flux values. Most of the flagged points occur in the vicinity of momentum dumps, when the spacecraft thrusters are engaged to reset the onboard reaction wheels. Momentum dumps occurred every 2--2.5 days and lasted up to about half an hour. Data taken during or near these momentum dumps typically display anomalous fluxes in the raw photometric time series. We also note that a large portion of data taken during Sector 3 observations suffered from poor pointing and other non-nominal instrumental behavior and were assigned NaN flux values by the SPOC pipeline; these points were removed, resulting in shorter segments of usable data in the two physical orbits of Sector 3.

\begin{figure*}
\centering
\includegraphics[width=0.9\linewidth]{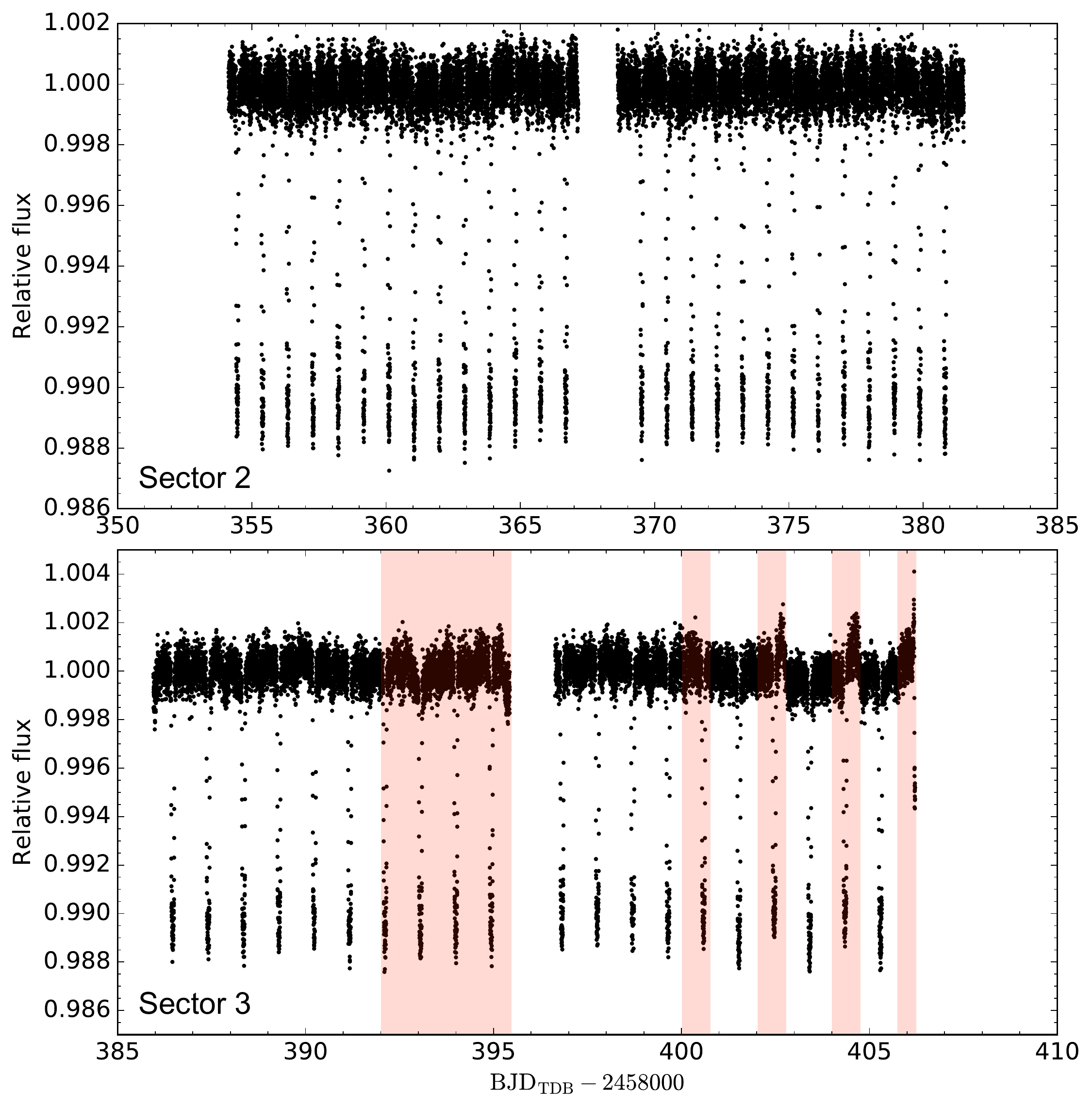}
\caption{Plot of the median-normalized and outlier-removed Presearch Data Conditioning (PDC) Sector 2 and 3 light curves of the WASP-18 system. The gaps in the middle of each sector's time series separate the two physical orbits of the \tess\ spacecraft contained within the sector. We have removed 30 minutes of data from the start of each orbit's light curve. The points highlighted in red show severe residual uncorrected systematics in the fits and are removed in the joint analysis presented in this paper. The phase curve variation of the system is clearly visible, as are residual long-term trends in the data. Those trends can include long-term photometric variability of the target or systematics introduced by the instrument and/or analysis.}
\label{fig:lightcurves}
\end{figure*}

We remove all flagged data points from the light curves. Then, we apply a moving median filter to the photometric time series with a width of 16 data points and remove \sig{3} outliers, while masking out the transits and secondary eclipses. We also remove the first 30 minutes of data from each orbit's light curve as these segments display residual ramp-like systematic artifacts. The resulting median-normalized light curves are shown in Figure~\ref{fig:lightcurves}. Even without detrending, the orbital phase variation is clearly discernible, as are the transits and, upon closer inspection, the secondary eclipses. The gaps in the middle of each sector's time series is due to the data downlink, separating the two physical orbits within each \tess\ Sector.

When fitting each orbit's light curve separately, several sections of the Sector 3 data display obvious features in the residuals that are not well-corrected by the systematics model we use (Section~\ref{subsec:systematics}). These sections occur before momentum dumps and at the end of an orbit's light curve. In the final light curve fits presented in this paper, we have carefully inspected the individual orbit light curve fits and removed regions with strong discernible uncorrected systematics. In the first orbit of Sector 3, we remove all points after BJD = 2,458,392, while in the second orbit, we remove 0.75~days worth of data before the last three momentum dumps and 0.5~days worth of data from the end of the time series. These regions are highlighted in red in Figure~\ref{fig:lightcurves}.

For Sector 2 data, trimming and outlier filtering remove 1.3\% and 1.6\% of the data from the two orbits, respectively. The additional removal of regions in Sector 3 data with uncorrected residuals entails a significantly higher percentage of removed points in the final Sector 3 light curves used in the joint analysis: 37\% and 29\% for the two orbits, respectively.

In addition to the phase curve modulation, there are clear residual long-term trends in the light curves at the level of several hundreds of parts per million (ppm). Previous analyses of \tess\ transit light curves have corrected for these and other flux variations by fitting a basis spline across the out-of-occultation (transit and secondary eclipse) light curve \citep{huang2018,vanderspek2019} or using a Gaussian Process (GP) model \citep{wang2019}, thereby removing all non-transit variability. Since such methods would remove the astrophysical phase variations of interest here, we do not utilize them and instead define a detrending model in our fits (Section~\ref{subsec:systematics}).

We have also carried out a parallel analysis using the Simple Aperture Photometry (SAP) light curves, which are not corrected for systematics by the SPOC pipeline. One particular characteristic of these data that is not manifested in the PDC light curves is significant flux ramps and periods of increased photometric scatter lasting up to a day at the start of each orbit's photometric time series and preceding each momentum dump. These features are not adequately detrended by the polynomial model we use in this work (Section~\ref{subsec:systematics}), so we choose to trim one day worth of data prior to each momentum dump, as well as the first day of data for each orbit. All in all, this removes almost 40\% of the SAP time series from the phase curve analysis. 

The best-fit astrophysical parameters from our analysis of SAP light curves are consistent with the results from the PDC light curves at much better than the \sig{1} level, with most parameters lying within 0.1-0.2$\sigma$. Meanwhile, the data trimming and larger red noise in the SAP light curves lead to parameter estimate uncertainties that are as much as 100\% larger than those derived from fitting the PDC light curves. Given the demonstrated consistency between the PDC and SAP light curve analyses and the poorer quality of the latter, we have decided to present the results from our PDC light curve analysis in this paper.

\section{Data Analysis}
\label{sec:dataanal}

In this work, we utilize the ExoTEP pipeline to analyze the \tess\ PDC light curves for the WASP-18 system. ExoTEP is a highly-modular Python-based tool in development for extracting and analyzing all types of time series photometric datasets of relevance in exoplanet science --- primary transit and secondary eclipse light curves and full-orbit phase curves. The pipeline allows the user to execute joint fits of datasets from multiple instruments (e.g., \textit{Kepler}, \textit{Hubble}, \textit{Spitzer}, and \tess) and customize the handling of limb-darkening and systematics models in a self-consistent way. So far, the main application of the ExoTEP pipeline has been in transmission spectroscopy \citep[e.g.,][see first reference for a detailed technical description]{wong2018,benneke2018,chachan2018}. This work is the first application of the ExoTEP pipeline to time series photometry that includes orbital phase curves.

\subsection{Transit and eclipse model}
\label{subsec:transit}

ExoTEP models both transits and secondary eclipses using the BATMAN package \citep{kreidberg2015}. In our analysis, we fit for the planet-star radius ratio $R_{p}/R_{s}$ and the relative brightness of the planet's day-side hemisphere $f_{p}$, which determine the planetary transit and secondary eclipse depths, respectively. In order to obtain updated values, we allow the transit geometry parameters --- impact parameter $b$ and scaled orbital semi-major axis $a/R_{s}$ --- to vary and fit for a new transit ephemeris --- specific mid-transit time $T_{0}$ and orbital period $P$. The zeroth epoch, to which we assign $T_{0}$, is designated to be the transit event closest to the center of the combined time series.

For the transit model, we use a standard quadratic limb-darkening law and fix the coefficients to values calculated for the \tess\ bandpass by \citet{claret2017}. Assuming the stellar properties for WASP-18 listed in \citet{stassun2017} and \citet{torres2012} ($T_{\mathrm{eff}}=6431\pm48$~K, $\log g=4.47\pm0.13$, $\mathrm{[Fe/H]}=0.11\pm0.08$; see also Table~\ref{tab:knownparams}), we take the coefficient values tabulated for the nearest-neighbor set of stellar properties ($T_{\mathrm{eff}}=6500$~K, $\log g=4.50$, $\mathrm{[Fe/H]}=0.1$): $u_{1}=0.2192$ and $u_{2}=0.3127$. 
Utilizing coefficients for other similar combinations of stellar properties does not significantly affect the results of our fit, yielding changes to the fitted parameter values of at most \sig{0.2}.
When experimenting with fitting for the quadratic limb-darkening coefficients, we obtain modestly constrained estimates that differ from the \citet{claret2017} coefficient values at the 3.1--3.4$\sigma$ level, while likewise maintaining the other fitted astrophysical parameter estimates at statistically consistent values. On the other hand, fitting for a single coefficient assuming a linear limb-darkening law yields estimates that differ significantly from the modeled values in \citet{claret2017} and introduces strong correlations between the limb-darkening coefficient and the transit depth and transit geometry parameters.

In the fits presented here, we fix the orbital eccentricity $e$ and argument of periastron $\omega$ to the values obtained by \citet{nymeyer2011}: $e=0.0091$ and $\omega=269^{\circ}$. For further discussion of orbital eccentricity, see Section~\ref{sec:res}.


\begin{deluxetable}{lllc}
\tablewidth{0pc}
\tabletypesize{\scriptsize}
\tablecaption{
    Known Parameters
    \label{tab:knownparams}
}
\tablehead{
    \multicolumn{1}{c}{~~~~~~~~Parameter~~~~~~~~} &
    \multicolumn{1}{c}{Value}                     &
    \multicolumn{1}{c}{Error}    &
    \multicolumn{1}{c}{Source}    
}
\startdata
$\teff$ (K)  \dotfill  & 6431 & 48 & \cite{stassun2017} \\ 
$\rstar$ (\rsun) \dotfill  & 1.26 & 0.04 & Stassun et~al., in prep.\tablenotemark{a}\\
$\mstar$ (\msun) \dotfill  & 1.46 & 0.29 & \cite{stassun2017} \\
$\log g$ \dotfill & 4.47 & 0.13 & \cite{stassun2017} \\
${\rm [Fe/H]}$ \dotfill & 0.11 & 0.08 & \cite{torres2012} \\
$K_{RV}$ (\ms)   \dotfill  & 1816.6 & $_{-6.3}^{+6.1}$ & \cite{knutson2014}  
\enddata
\tablenotetext{a}{Stellar radius is derived using Gaia DR2 data.}
\end{deluxetable}

\subsection{Phase curve model}
\label{subsec:phase}

We model the out-of-occultation variation of the system brightness as a third-order harmonic series in phase \citep[e.g.,][]{carter2011}:
\begin{equation}\label{eq:prefull}
\psi'(t) = 1+\bar{f_{p}}+\sum\limits_{k=1}^{3}A_{k}\sin(k\phi(t))+\sum\limits_{k=1}^{3}B_{k}\cos(k\phi(t)).\end{equation}
Here, we have normalized the flux such that the average brightness of the star alone is unity. The baseline relative planetary brightness $\bar{f_{p}}$ is the average of the planet's apparent flux across its orbit. The phase function $\phi(t)$ is derived from the time series via the relation $\phi(t) = 2\pi(t-T_{0})/P$, where $t$ is time.

Several of the harmonic terms in the phase curve model are attributed to various physical processes on the star or planet. The star's brightness is modulated by the beaming effect and ellipsoidal variation. These two processes produce phase curve signals at the fundamental of the sine ($A_{1}$) and the first harmonic of the cosine ($B_{2}$), respectively. A tidally-locked hot Jupiter has a fixed day-side hemisphere facing the star, which produces a variation in its apparent brightness due to the changing viewing geometry. This atmospheric brightness component produces a signal at the fundamental of the cosine ($B_{1}$). See the discussion in Section~\ref{sec:dis} for more details concerning the astrophysical implications of the phase curve terms.

We can separate the total system phase curve model into terms describing the star's brightness variation $\psi_{*}(t)$ and terms describing the planet's brightness variation $\psi_{p}(t)$:
\begin{align}
\psi_{p}(t) & = f_{p}-|B_{1}|+B_{1}\cos(\phi(t)), \\
\psi_{*}(t) & = 1 + \sum\limits_{k=1}^{3}A_{k}\sin(k\phi(t))+\sum\limits_{k=2}^{3}B_{k}\cos(k\phi(t)).
\end{align}
We have assigned all terms without a direct corresponding physical process to the star's phase modulation. These other terms can become significant if there are discernible phase shifts in the aforementioned modulation signals. We have also used the fact that the average brightness of the planet is the maximum measured brightness, which occurs at secondary eclipse, minus the semi-amplitude of the atmospheric variation: $\bar{f_{p}}\equiv f_{p}-|B_{1}|$. 

The separation of stellar and planetary phase curve terms is important when including the transit and eclipse light curves $\lambda_{t}(t)$ and $\lambda_{e}(t)$, since only the brightness modulation of the occulted region on the star or planet is removed from the total system flux. This correction is particularly consequential during secondary eclipse, when the atmospheric brightness component is completely blocked by the star ($\lambda_{e}=0$), while the ellipsoidal and beaming modulations on the star are unaffected ($\lambda_{t}=1$). From here, we can write down the full phase curve model, including eclipses (transit and secondary eclipse) and re-normalized such that the average out-of-eclipses flux is unity:
\begin{equation}\label{full}
\psi(t) = \frac{\lambda_{t}(t)\psi_{*}(t)+\lambda_{e}(t)\psi_{p}(t)}{1+f_{p}-|B_{1}|}.
\end{equation}

When compared to the standard approach in the literature, which simply multiplies the system phase curve model in \eqr{prefull} and the occultation light curves together \citep[e.g.,][]{carter2011}, this more detailed and physical model deviates most significantly during the ingress and egress of secondary eclipse, where the discrepancy for the WASP-18 system can be as large as several tens of ppm. While this level of model discrepancy is well within the noise of the light curves analyzed in this work, studies with higher signal-to-noise photometry would benefit from our more careful treatment of the phase curve model.

\subsection{Modeling long-term trends}
\label{subsec:systematics}

As can be seen in Figure~\ref{fig:lightcurves}, the \tess\ PDC light curves from the SPOC pipeline show long-term trends at the level of several hundreds of ppm. Those can be caused by low-frequency residual systematics and/or long-term stellar variability. 
 
For the detrending model, we use a polynomial in time of the form
\begin{equation} 
S^{\lbrace i\rbrace}_{n}(t) = \sum\limits_{k=0}^{n}c^{\lbrace i\rbrace}_{k}(t-t_{0})^{k},
\end{equation}
where $t_{0}$ is the first time stamp in the light curve from orbit $i$, and $n$ is the order of the polynomial model. In the following, we assign $i=1,2$ to the two orbits of the \tess\ spacecraft that comprise the full Sector 2 data (those correspond to physical orbits 11 and 12, where the numbering started during commissioning) and $i=3,4$ to the two orbits contained in Sector 3 data (corresponding to physical orbits 13 and 14). The complete phase curve model is therefore
\begin{equation}
f(t) = S^{\lbrace i\rbrace}_{n}(t)|_{i=1,2,3,4} \times \psi(t).
\end{equation}

To determine the optimal polynomial order for each orbit, we carry out phase curve fits of individual orbit light curves and choose the order that minimizes the Bayesian Information Criterion (BIC), which is defined as $\mathrm{BIC} \equiv k\log n-2\log L$, where $k$ is the number of free parameters in the fit, $n$ is the number of data points, and $L$ is the maximum log-likelihood.

For the first orbit, we find that a 7th-order polynomial model minimizes the BIC, while for the second orbit, we use a 9th-order polynomial. For the two shorter Sector 3 orbits light curves, we use a 5th-order and 3rd-order polynomial, respectively. 

The use of a time-dependent detrending model when fitting for a time-varying astrophysical signal can sometimes introduce artificial biases into the parameter estimates. When selecting polynomials of similar order, the best-fit astrophysical parameters do not vary by more than 0.2$\sigma$. To further assess the effects of our use of polynomial detrending, we carry out a special joint fit of all four orbital light curves without using any detrending model, instead using a simple multiplicative normalization factor for each orbit's light curve. The resultant estimates of the transit/eclipse depths and phase curve amplitudes are consistent with the results of the full joint fit with detrending models (\secr{res} and \tabr{modelparams}) at better than the \sig{1} level. This test demonstrates that the results presented in this paper are highly robust to the particular choice of detrending model.

\subsection{Contaminating sources}
\label{subsec:contamination}

Since the primary objective of the \tess\ mission is to carry out a survey of nearby, bright stars in search of transiting planets, the camera focus is set so as to spread a point source's flux over several pixels to adequately sample the point spread function and achieve high photometric precision. As a benchmark, approximately 50\% and 90\% of a star's flux is contained within a 1$\times$1 and 4$\times$4~pixel region around the centroid, respectively \citep{ricker2015}. Given the pixel scale of $21''$, this indicates that a star's pixel response function (PRF) occupies a significant on-sky area, raising the possibility of contaminating sources overlapping with the target PRF and blending with the extracted photometry.

There are two moderately bright stars within the \tess\ input catalog that are in the vicinity of WASP-18 --- TIC 100100823 and TIC 100100829. These two nearby sources lie $73''$ and $83''$ away from the target and have \tess\ magnitudes of $T=12.65$ mag and $T=12.50$ mag, respectively. When compared to WASP-18 ($T=8.83$ mag) the nearby sources are 34 and 29 times fainter, respectively.

The optimal apertures selected by the SPOC pipeline for photometric extraction are shown in Figure~\ref{fig:stamp}. The locations of the two nearby stars are also indicated. The SPOC pipeline uses a model PRF derived from commissioning data to remove the flux from neighboring sources located on each target's subarray. The relative uncertainty of the deblending process due to imperfections in the PRF model and intrinsic variations in the PRFs of different sources across the detector is estimated to be at the level of a few percent \citep{jenkins2010}. 

In the Sector 2 pixel stamp, TIC 100100823 is 0.66~pixels away from the edge of the aperture at its closest point. Using the aforementioned benchmark estimates of a point source's flux distribution on the detector, we determine that the central 50\% of its undeblended flux (from the 1$\times$1~pixel region around its centroid) was not included in the aperture. Another 40\% of its flux was distributed in the remaining 15 pixels in the surrounding 4$\times$4~pixel region. Tracing a 4$\times$4~pixel region centered on the location of TIC 100100823, with edges aligned with the pixel boundaries, we can estimate that the fractional area lying within the extraction aperture is $\sim$0.25. Assuming that the flux was evenly distributed across the 15-pixel region to obtain a generous upper limit, we predict that at most $0.25\times 40\% = 10\%$ of the star's flux lay within the science aperture prior to deblending. Multiplying this upper limit with the estimated uncertainty in the deblending process and the flux of the star relative to WASP-18 gives a relative deblending contamination contribution of roughly 0.01\%. The other similarly-bright nearby star, TIC 100100829, is further away from the edge of the aperture, so its level of residual contamination in the extracted photometry is smaller. For Sector 3, the nearby stars are situated farther away from the edge of the optimal aperture, and thus the expected deblending contamination contribution is even more negligible.

As detailed in the following section, the relative uncertainties we obtain for the fitted astrophysical parameters that would be directly affected by such contamination  --- phase curve amplitudes and transit/eclipse depth --- lie above the 0.13\% level (in fact, for all parameters except for transit depth, the relative uncertainties are greater than 3\%). This means that any effect stemming from uncorrected contamination is overshadowed by the much larger intrinsic uncertainties in the parameter estimates, given the sensitivity of the data. Therefore, we do not consider contamination in our phase curve analysis.

\begin{deluxetable}{lll}
\tablewidth{0pc}
\tabletypesize{\scriptsize}
\tablecaption{
    Model Parameters
    \label{tab:modelparams}
}
\tablehead{
    \multicolumn{1}{c}{~~~~~~~~~~Parameter~~~~~~~~~~} &
    \multicolumn{1}{l}{Value}                     &
    \multicolumn{1}{l}{Error}    
}
\startdata
\sidehead{\textit{Fixed Parameters}}
$e$\tablenotemark{a}   \dotfill  & 0.0091 & --- \\
$\omega$\tablenotemark{a} ($^{\circ}$) \dotfill  & 269 & --- \\
$u_{1}$\tablenotemark{b} \dotfill & 0.2192 & --- \\
$u_{2}$\tablenotemark{b} \dotfill & 0.3127 & --- \\ 

\sidehead{\textit{Fitted Parameters}}
$R_p/R_s$   \dotfill  & 0.09716 & $_{-0.00013}^{+0.00014}$ \\
$f_p$ (ppm) \dotfill  & 341 & $_{-18}^{+17}$ \\
$T_0$ (BJD$_{\mathrm{TDB}}$)      \dotfill  & 2458375.169883 & $_{-0.000025}^{+0.000026}$ \\
$P$ (days)  \dotfill  & 0.9414526 & $_{-0.0000015}^{+0.0000016}$\\
$b$         \dotfill  & 0.318 & $_{-0.019}^{+0.018}$\\
$a/R_s$     \dotfill  & 3.562 & $_{-0.023}^{+0.022}$\\
$A_1$ (ppm) \dotfill  & 21.0  & 4.5 \\
$A_2$ (ppm) \dotfill  & -4.0   & 4.6 \\
$A_3$ (ppm) \dotfill  & -14.0  & 4.6 \\
$B_1$ (ppm) \dotfill  & -174.4 & $_{-6.2}^{+6.4}$ \\
$B_2$ (ppm) \dotfill  & -190.5 & $_{-5.9}^{+5.8}$ \\
$B_3$ (ppm) \dotfill  & -3.9  & 6.1\\
$\sigma_{1}$ (ppm) \dotfill  & 527.6 & $_{-4.1}^{+3.9}$ \\
$\sigma_{2}$ (ppm) \dotfill  & 528.6 &  3.8\\
$\sigma_{3}$ (ppm) \dotfill  & 526.5 &  5.6\\
$\sigma_{4}$ (ppm) \dotfill  & 521.4 &  $_{-5.2}^{+5.3}$\\
\sidehead{\textit{Derived Parameters}} 
Transit depth\tablenotemark{c} (ppm)  \dotfill  & 9439 & $_{-26}^{+27}$\\
$i$ ($^{\circ}$)    \dotfill  & 84.88 & 0.33 \\
$R_p$ ($R_{\mathrm{Jup}}$)\dotfill & 1.191 & 0.038\\
$a$ (AU)\dotfill & 0.02087 & 0.00068
\enddata
\tablenotetext{a}{Best-fit values from \citet{nymeyer2011}.}
\tablenotetext{b}{Tabulated in \citet{claret2017}.}
\tablenotetext{c}{Calculated as $(R_p/R_s)^2$.}
\end{deluxetable}

\begin{figure*}
\includegraphics[width=\linewidth]{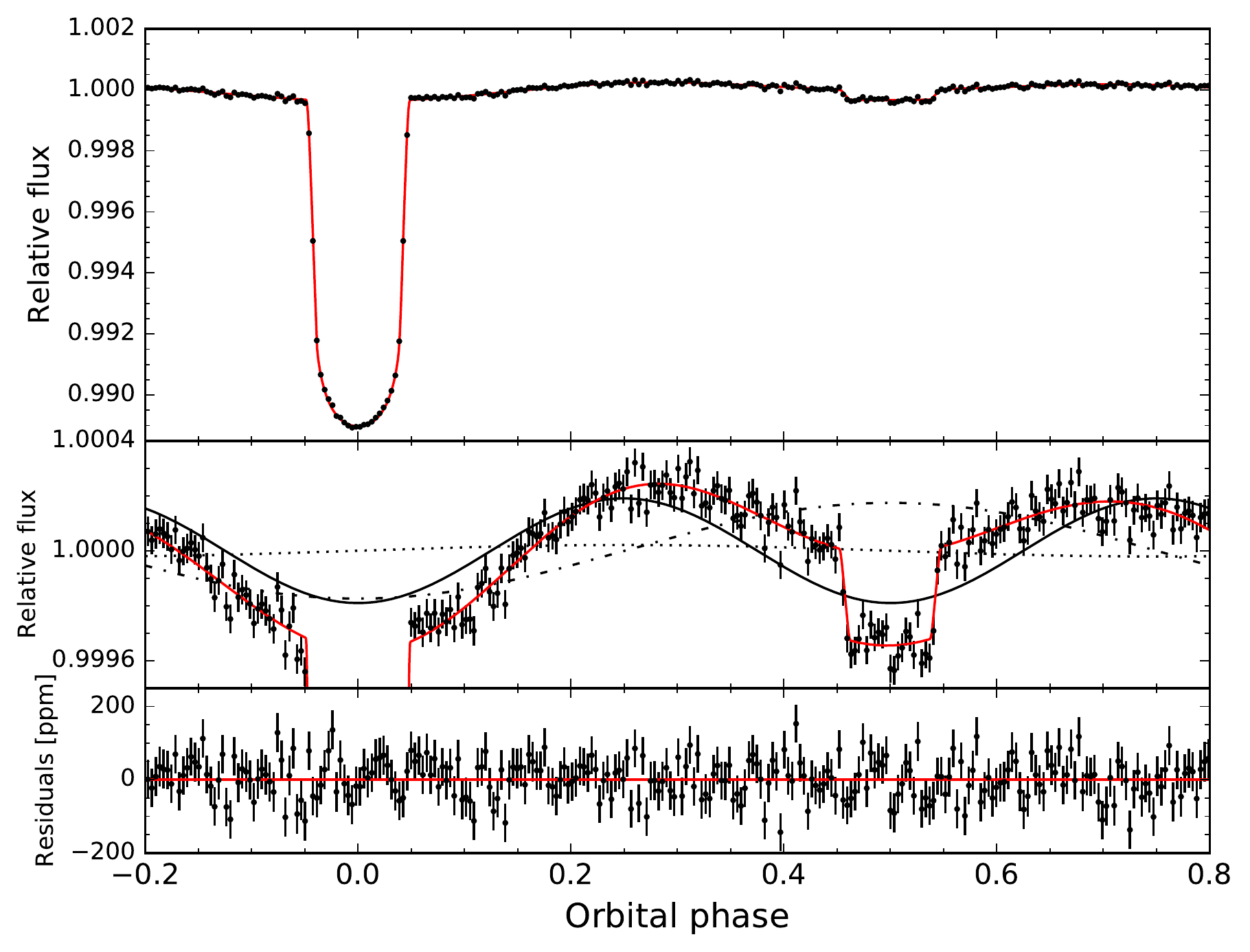}
\caption{Top panel: The phase-folded light curve, after correcting for long-term trends, binned in 5-minute intervals (black points), along with the best-fit full phase curve model from our joint analysis (red line). Middle panel: same as top panel, but with an expanded vertical axis to detail the fitted phase curve modulation and secondary eclipse signal. The contributions to the phase curve from ellipsoidal ($B_{2}$), atmospheric brightness ($B_{1}$), and beaming ($A_{1}$) modulations are plotted individually with black solid, dot-dash, and dotted lines. Bottom panel: plot of the corresponding residuals from the best-fit model.}
\label{fig:bestfit}
\end{figure*}

\subsection{Model fit}
\label{sec:fit}

We carry out a joint fit of the WASP-18 light curves from Sectors 2 and 3. We fit for $R_p/R_s$, $f_{p}$, $T_{0}$, $P$, $b$, $a/R_{*}$, $A_{1-3}$, $B_{1-3}$, $c^{\lbrace 1\rbrace}_{0-7}$, $c^{\lbrace 2\rbrace}_{0-9}$, $c^{\lbrace 3\rbrace}_{0-5}$, and $c^{\lbrace 4\rbrace}_{0-3}$. In addition, we fit for a uniform per-point uncertainty for each orbit --- $\sigma_{1-4}$ --- which ensures that the resultant reduced $\chi^2$ value is near unity and self-consistently produces realistic uncertainties on the other parameters given the intrinsic scatter of the light curves. The total number of free astrophysical, long-term trend, and noise parameters is 44. 

The ExoTEP pipeline simultaneously computes the best-fit values and posterior distributions for all parameters using the affine-invariant Markov Chain Monte Carlo (MCMC) ensemble sampler \texttt{emcee} \citep{emcee}. We set the number of walkers to four times the number of free parameters: 176. To facilitate the convergence of the chains we initialize the walkers with values close to those of the best fit to light curves of individual physical orbits. The length of each walker's chain is 100,000 steps, and we discard the first 80\% of each chain before calculating posterior distributions. The chains are plotted and visually inspected to ensure convergence. We also re-run the entire process, to confirm that the resultant parameter estimates are consistent to well within $0.1\sigma$.

\section{Results}
\label{sec:res}

The results of our joint analysis are listed in Table~\ref{tab:modelparams}, where we include astrophysical and noise parameters. The median and \sig{1} uncertainties derived from the posterior distributions are given. The combined phased light curve with long-term trends removed is plotted in Figure~\ref{fig:bestfit} along with the best-fit full phase curve model.

Comparing with other published values in the literature \citep{hellier2009, southworth2009, nymeyer2011}, we find that our results generally lie in good agreement with previous estimates. The calculated values of the orbital parameters $i$ and $a/R_{*}$ are the most precise to date and are well within \sig{1} of the values presented in the discovery paper \citep{hellier2009} and follow-up photometric studies \citep{southworth2009}. We obtain an updated and refined mid-transit time with a precision of 2.2~seconds and a period estimate that is consistent at better than the \sig{0.2} level with the results of \citet{hellier2009} as well as recent studies that fitted ephemerides across all previously measured transit times \citep{wilkins2017,mcdonald2018}.

The measured transit depth $(R_{p}/R_{*})^{2}$ of $9439_{-26}^{+27}$~ppm is significantly (\sig{3.3}) larger than the estimate of \citet{hellier2009} of $8750\pm 210$~ppm. A separate analysis of the photometry and radial velocity (RV) data obtained by \citet{hellier2009} incorporating additional RV measurements yielded a somewhat larger transit depth estimate of $9160^{+200}_{-120}$~ppm \citep{triaud2010}, which is more consistent (\sig{1.4}) with the value derived here. \citet{southworth2009} carried out a thorough analysis of WASP-18b transit light curves using a variety of limb-darkening laws and fitting methodologies and obtained a wide range of transit depths (9300--9800~ppm) consistent with the estimate from our analysis.

The most notable results from our phase curve analysis are the $19\sigma$ detection of a secondary eclipse in the \tess\ bandpass and the robust detection of phase curve variations corresponding to beaming ($A_{1}$), atmospheric brightness ($B_{1}$), and ellipsoidal ($B_{2}$) modulations (See table~\ref{tab:modelparams}). These three leading phase curve harmonics are plotted individually in the middle panel of Figure~\ref{fig:bestfit}.

In our analysis, we have fixed the orbital eccentricity $e$ and argument of periastron $\omega$ to the most recent literature values \citep{nymeyer2011}. Since the orbit is very nearly circular, we have experimented with carrying out fits where we fix $e=0$, as well as fits where we allow $e$ and $\omega$ to vary freely. In both cases, we obtain parameter estimates that are consistent with the best-fit values listed in Table~\ref{tab:modelparams} at better than the $0.9\sigma$ level. 

In the free-eccentricity fits, we obtain relatively weak eccentricity constraints: $e=0.0015_{-0.0011}^{+0.0038}$ and $\omega=156_{-69}^{+115}$~deg. These estimates are consistent with those published in \citet{nymeyer2011} at better than the \sig{2} level. In their analysis of high signal-to-noise thermal infrared secondary eclipse light curves from \textit{Spitzer}, they obtain orbital eccentricity estimates that are much more precise than what we can constrain from our analysis. Therefore, we have decided to fix $e$ and $\omega$ to their best-fit values in the joint fits described in Section~\ref{sec:res}.

The fitted per-point uncertainties for the light curves $\sigma_{1-4}$ are nearly identical at 520--530~ppm, indicating that the noise level in the data is consistent across the four orbits. The standard deviations of the residuals binned at 1-hour intervals in time are 130, 143, 142, and 158~ppm for the four orbits, respectively, which are generally consistent with the benchmark prediction of 123~ppm calculated for a $T=8.83$~mag target based on simulated photometry for the \tess\ mission \citep{sullivan2015,stassun2017}. The larger binned residuals in the Sector 3 light curves (particularly in the last orbit) indicate suboptimal photometric performance, as discussed in \secr{obs}.

\section{Discussion}
\label{sec:dis}

Among the sinusoidal phase curve coefficients fitted in our model, only three have a high statistical significance (with a signal to noise ratio of well over 3, see \tabr{modelparams}) --- those corresponding to the fundamental, $A_1$ and $B_1$, and the first harmonic of the cosine, $B_2$. As discussed in Section~\ref{subsec:phase}, these terms are attributed to the beaming, atmospheric, and ellipsoidal components, respectively. Each component is discussed in Sections~\ref{sec:beam}--\ref{sec:atm} below. 
 
For the beaming and ellipsoidal components we show in Sections~\ref{sec:beam} and \ref{sec:tide} that their amplitudes and signs are consistent with theoretical expectations. While that consistency can be expected, there are systems where that is not the case for one of those phase components and cases where the mass ratios derived from the two amplitudes do not agree \citep[e.g.,][see also \citealt{shporer2017} Section 3.4]{vankerkwijk2010, carter2011, shporer2011, barclay2012, bloemen2012, esteves2013, faigler2015, rappaport2015}. One possible explanation for that disagreement is a phase shift in the atmospheric component \citep[e.g.,][]{faigler2015, shporer2015}. Another possible explanation is poor modeling of the tidal distortions of hot stars, given their fast rotation and lack of convective zone in their atmospheres \citep{pfahl2008}.

Our phase curve model includes three other components --- the first harmonic of the sine, $A_2$, and the second harmonic terms, $A_3$ and $B_3$ (see \tabr{modelparams}).
The first harmonic of the sine, $A_2$, is not statistically significant which is consistent with theoretical expectations as we are not aware of any astrophysical process associated with that coefficient \citep[e.g.,][]{faigler2011, shporer2017}.  
The second harmonic of the cosine ($B_3$) is expected as a higher-order term of the ellipsoidal distortion modulation, and the fitted amplitude, although not statistically significant, is in agreement with expectations (see more details in \secr{tide}).
Finally, the sine amplitude of the second harmonic, $A_3$, is $-14.0 \pm 4.6$ ppm, at the \sig{3} level. While it is smaller in amplitude and lower in statistical significance than the coefficients associated with the beaming, ellipsoidal, and atmospheric components ($A_1$, $B_2$, and $B_1$, respectively, see \tabr{modelparams}), the astrophysical origin of that coefficient is not immediately clear. It is interesting to note that a similar phase component was measured at a statistically significant level for Kepler-13Ab \citep{esteves2013, shporer2014}, although that host star is a hot A-type star \citep{shporer2014} while WASP-18 is a mid F-type star.

We have experimented with fitting a simplified phase curve model without the second harmonic terms, i.e., including only the fundamental ($A_{1}$, $B_{1}$) and first harmonic ($A_{2}$, $B_{2}$) terms. We do not find that any of the fitted astrophysical parameter estimates change by more than $0.3\sigma$. This indicates that our phase curve analysis is not significantly affected by the inclusion of the second harmonic terms.

\subsection{Beaming}
\label{sec:beam}

The sine component of the fundamental, $A_1$, is expected to result from the beaming modulation. We derive the expected amplitude using:
\begin{equation}
    A_{\rm beam} = \alpha_{\rm beam} 4 \frac{K}{c} \ ,
    \label{eq:beam}
\end{equation}
where $K$ is the orbital RV semi-amplitude, and $c$ the speed of light. The $\alpha_{\rm beam}$ coefficient is of order unity and depends on the target's spectrum in the observed bandpass (for a more detailed description of the nature of this coefficient, see \citealt{shporer2017}). 

Assuming the target is a blackbody and integrating across the \tess\ bandpass, we derive an expected beaming modulation amplitude of $A_{\rm beam}$ = $18 \pm 2$~ppm, using the known parameters of the system (see \tabr{knownparams}). This is consistent with the measured value of $21.0 \pm 4.5$~ppm. Examining the shape of the phase curve in Figure~\ref{fig:bestfit}, the presence of a significant beaming modulation signal is seen as the difference between the two brightness maxima within the orbit.

\subsection{Tidal interaction}
\label{sec:tide}


The ellipsoidal modulation measured in the phase curve is the result of tidal interaction between the planet and the star. To estimate the expected value of that photometric modulation amplitude, we use the following approximate equation \citep{morris1985, morris1993}: 
\begin{equation}
    A_{\rm ellip} = \alpha_{\rm ellip} \frac{M_p \sin^2 i}{M_s}\left(\frac{R_s}{a}\right)^3  \ \ ,
    \label{eq:aellip}
\end{equation}
    where $M_s$ and $R_s$ are the host star mass and radius, respectively, $M_p$ the planet mass, $a$ the orbital semi-major axis, and $i$ the orbital inclination angle. The $\alpha_{\rm ellip}$ coefficient is derived from the linear limb darkening coefficient $u$ and gravity darkening coefficient $g$:
\begin{equation}
    \alpha_{\rm ellip} = 0.15 \frac{(15+u)(1+g)}{3-u} \  \ .
\end{equation}
We estimate $u$ and $g$ using the known parameters of the host star (see \tabr{knownparams}) and the tables of \cite{claret2017} to arrive at $\alpha_{\rm ellip} = 1.10 \pm 0.05$. This gives an expected amplitude of $A_{\rm ellip} = 172.5 \pm 14.6$~ppm (uncertainty derived from uncertainty of known parameters), in good agreement with the measured amplitude of $|B_2|=190.5_{-5.8}^{+5.9}$ ppm.

As noted earlier, \eqr{aellip} is an approximation, as it is the leading term in a Fourier series \citep{morris1985, morris1993}. We have calculated the next term in the series, the coefficient of the cosine of the 2nd harmonic of the orbital period \citep{morris1985}, to be $-11.8 \pm 1.8$ ppm. This is in agreement with the measured amplitude of $B_3 = -3.9 \pm 6.1$ ppm (see \tabr{modelparams}), although the latter is not statistically significant.

\subsection{Atmospheric characterization}
\label{sec:atm}

The drop in flux during secondary eclipse, as the planet is occulted by its host star, is due to the blocking of (1) starlight reflected by the planet's atmosphere and (2) thermal emission from the planet's atmosphere, since, given the strong stellar irradiation, the planet's thermal emission is expected to be significant at visible wavelengths. To estimate the planet's thermal emission in the \tess\ band, we use the atmospheric model of \cite{arcangeli2018}, derived by fitting to the measured secondary eclipse depths in the four Spitzer/IRAC bands (centered at 3.6, 4.5, 5.8, and 8.0~\mic) and HST/WFC3 (1.1--1.7~\mic). Integrating that atmospheric model across the \tess\ bandpass gives an expected thermal emission of 327~ppm. It is difficult to accurately quantify the uncertainty on the expected thermal emission and we estimate it to be at the few percents level (the uncertainty on the star-planet radii ratio derived here, which contributes to the thermal emission uncertainty, is only 0.14\%).

We measure a secondary eclipse depth of $341_{-18}^{+17}$~ppm. While this depth is significantly shallower than the $z'$ band eclipse depth of $682 \pm 99$ ppm measured by \cite{kedziora2019}, it is consistent with the expected thermal emission predicted by the atmospheric models of \cite{arcangeli2018}.

Since the difference between the measured secondary eclipse depth and the expected thermal emission in the \tess\ bandpass is only at a \sig{0.8} significance, we cannot claim a detection of reflected light, and therefore place a \sig{2} upper limit of 35~ppm on the relative contribution of reflected light to the atmospheric brightness. This upper limit is consistent with the upper limit on reflected polarized light from the planet measured by \cite{bott2018}.

The upper limit on the reflected light translates to an upper limit on the geometric albedo in the \tess\ bandpass of $A_g <$ 0.048 (\sig{2}), using the model parameters measured here (specifically $R_p/R_s$ and $a/R_s$). This upper limit is consistent with low visible-light geometric albedo measured for other short period gas giant planets, like WASP-12b ($A_g < $ 0.064 at 97.5\% confidence, \citealt{bell2017}), HD~209458b ($A_g = 0.038 \pm 0.045$, \citealt{rowe2008}), TrES-2b ($A_g = 0.0253 \pm 0.0072$, \citealt{kipping2011}), Qatar-2b ($A_g < 0.06$ at \sig{2}, \citealt{dai2017}), and others \citep[e.g.,][]{heng2013, esteves2015, angerhausen2015}.

The small geometric albedo suggests that the bond albedo is also small since the \tess\ bandpass is close to the wavelength region where the host star is brightest. The low albedo is consistent with a correlation between decreasing albedo and increasing planet mass, suggested by \citet{zhang2018}, although there is currently no theoretical mechanism to explain such a correlation.


The measurements of the secondary eclipse depth and the atmospheric phase component amplitude allow us to estimate the flux from the planet's night side --- the hemisphere facing away from the star, visible to the observer during transit. The night-side flux is the difference between the secondary eclipse depth $f_p$ and the full amplitude of the atmospheric brightness modulation $2 \times |B_{1}|$, which is $-8\pm 22$ ppm (see \tabr{modelparams}). Therefore, we do not measure statistically significant flux from the planet's night side, and place a \sig{2} upper limit of 43~ppm on the night side's brightness in the \tess\ bandpass, which is 13\% of the day-side brightness. Measurements of the orbital phase curve in the near-infrared at 3.6 and 4.5~\mic\ were also unable to detect flux from the night side \citep{maxted2013}. This points to very low efficiency of longitudinal heat distribution from the day side to the night side, as suggested by other authors \citep{nymeyer2011,maxted2013} and consistent with similar findings for other highly irradiated hot Jupiters \citep[e.g.,][]{wong2015,wong2016} and theoretical models \citep{perez2013, komacek2017}.

We have assumed here that the atmospheric phase variability is a sinusoidal modulation, where the maximum is coincident with the phase of secondary eclipse, and the minimum occurs at mid-transit. Deviations from this simplistic model have been observed, where the planet's surface brightness distribution is such that the brightest region is shifted away from the sub-stellar point, leading to a shift between the phase of maximum light and the center of secondary eclipse \citep[e.g.,][]{demory2013, shporer2015, hu2015, faigler2015, parmentier2016}. In our phase curve analysis a phase shift of the atmospheric phase component will manifest itself as a deviation of the beaming phase component amplitude from the theoretically predicted value, since the beaming and atmospheric components are the sine and cosine of the fundamental (the orbital period). The beaming amplitude is consistent with the predicted amplitude (see \secr{beam}), and we therefore place an upper limit on a phase shift of 2.9~deg (\sig{2}), derived from the fitted values and uncertainties of $A_1$ and $B_1$ (see \tabr{modelparams}). 

In principle, since the observed modulation is a superposition of modulations due to reflected light and thermal emission, each of the two processes may have a phase shift that is canceled out in the combined light. However, phase curves in the near-infrared do not show a phase shift of the thermal emission, down to 5--10~deg \citep{maxted2013}. Hence, we can rule out a phase shift of the reflected light modulation in the optical. The lack of a phase shift is consistent with the inefficient heat redistribution from the day-side to the night-side hemisphere.


\subsection{Detecting non-transiting WASP-18b-like objects}
\label{sec:nontransit}

The clear sinusoidal modulations seen along the orbital phase (see \figr{lightcurves}) suggests that they could be identified even if the system were not in a transiting configuration, i.e., with a smaller orbital inclination angle. If so, this in turn raises the possibility of detecting non-transiting but otherwise similar systems, as explored by \citet[][see also \citealt{faigler2012,faigler2013,talor2015, millholland2017}]{faigler2011} using a phase curve analysis.

\begin{figure}
\begin{centering}
\includegraphics[width=0.4\textwidth]{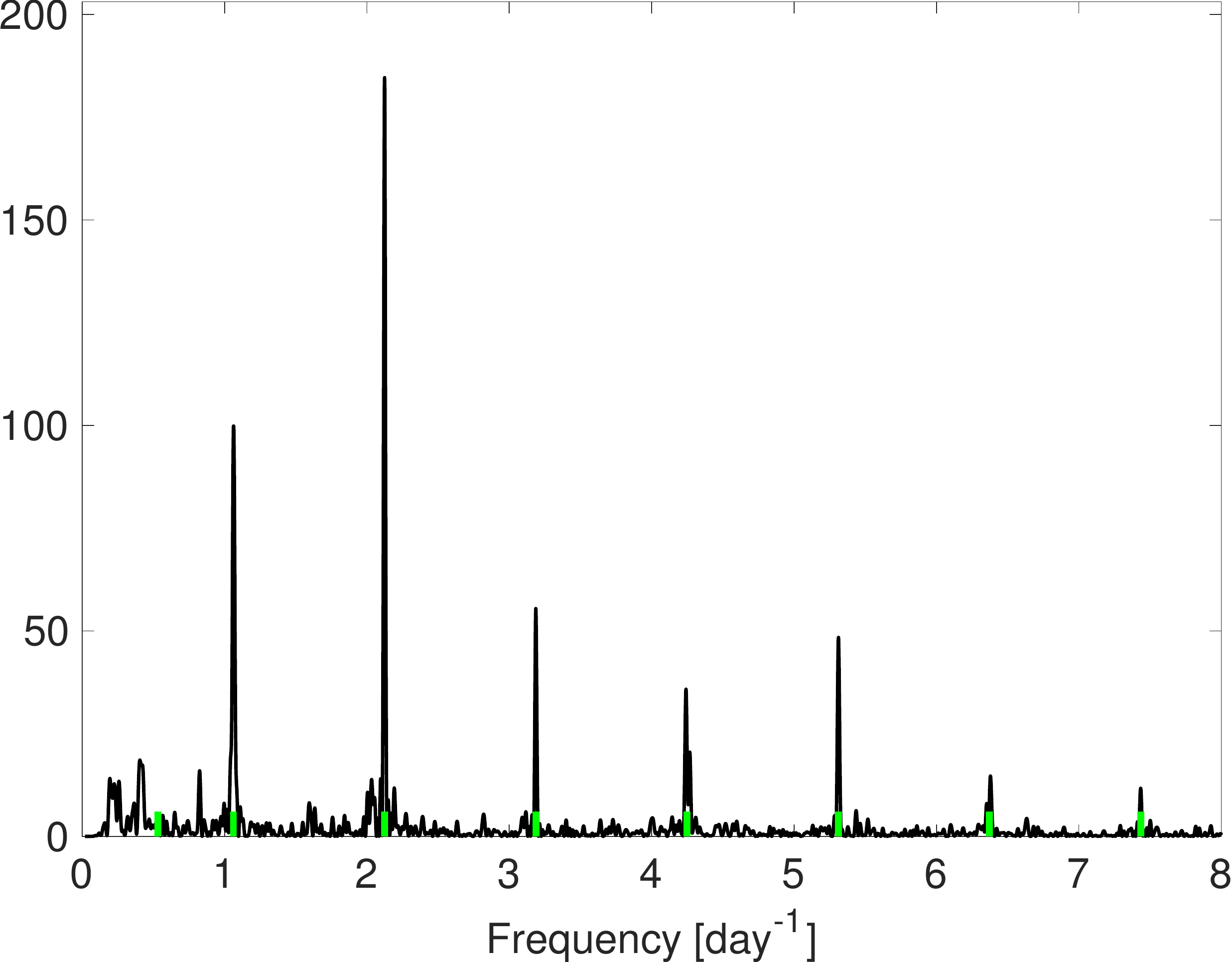}
\includegraphics[width=0.4\textwidth]{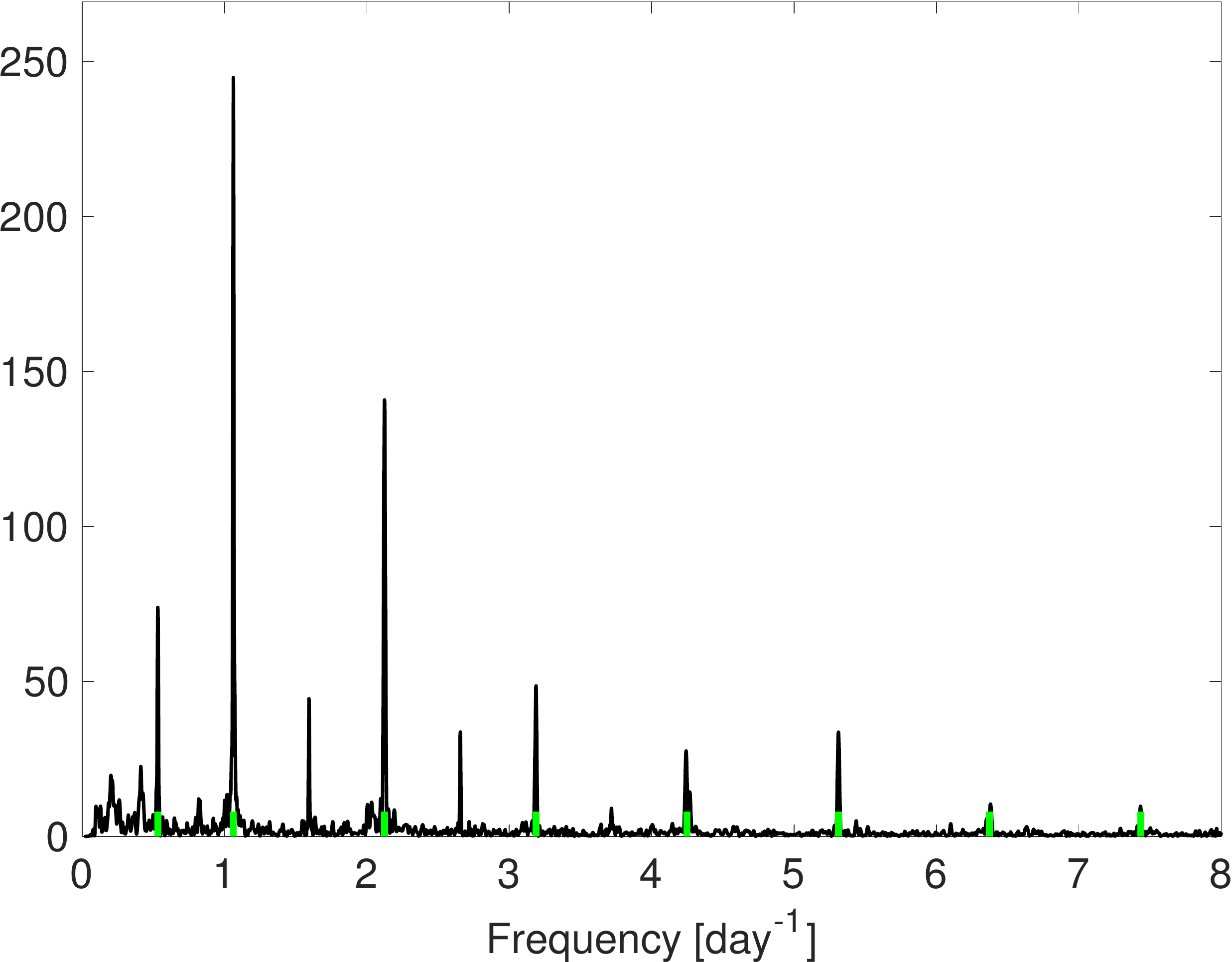}
\caption{Lomb-Scargle (top) and double-harmonic (bottom) periodograms of the lightcurves where data within the transits and secondary eclipse were removed, and normalized by their standard deviation (estimated through the median absolute deviation, or MAD). Both periodograms are dominated by the orbital frequency and its harmonics and sub-harmonic, marked by short vertical green lines on the x-axis. 
\label{fig:ls}}
\end{centering}
\end{figure}

To test that possibility, we have removed all data points within the transit and secondary eclipse and have carried out a period analysis of the remaining light curve, similar to the analysis done in \citet[][see also \citealt{shporer2014}]{shporer2011} for the Kepler-13 system. \figr{ls} shows both the Lomb-Scargle periodogram \citep{lomb1976, scargle1982} and the double-harmonic periodogram (following \citealt{faigler2011}). Both periodograms are dominated by variability at the orbital period. This demonstrates that variability at the orbital period can be clearly detected for non-transiting but otherwise similar systems. This also further confirms the prediction of \citet[][see their Figure 10]{shporer2017} that the orbital phase curve variability of systems such as WASP-18, containing massive planets on short periods, can be detected in \tess\ data.

It is important to note that the detection of periodic variability does not uniquely reveal the nature of the system as a star-planet system, since similar variability can be induced by a stellar-mass companion, stellar pulsations, and/or the combination of stellar activity (in the form of starspots) and stellar rotation. However, a star rotating at a period as short as the WASP-18b orbital period would show large rotational broadening of the spectral lines, with a rotation velocity over 50~\kms, which can be measured with a single stellar spectrum. Furthermore, as shown by \citet{faigler2011}, the measured amplitudes can be compared with the expected amplitudes based on the stellar parameters.

\subsection{TTV analysis}
\label{sec:ttv}

To complement our analysis of the phase curve we have also analyzed the individual transit times in order to look for transit timing variations (TTV). That analysis was divided into long term TTV, using all available transit times in the literature spanning about a decade, and short term TTV, within the time scale of the \tess\ Sector 2 and 3 data. Both analyses are described below.

\subsubsection{Long term TTV}
\label{sec:ttvlong}

Most hot Jupiters are vulnerable to tidal orbital decay, and their
orbits should be shrinking \citep[e.g.,][]{levrard2009,matsumura2010}. However, directly observing tidal inspiral has proven an obstinate challenge; there have not yet been any unambiguous detections of orbital decay due to tides (see \citealt{maciejewski2016,patra2017,bailey2019}
for discussion of the most promising case yet).
WASP-18b is a promising target in the search for tidal orbital decay.
Compared to other hot Jupiters, it has a particularly high planet to
star mass ratio, and an exceptionally small separation from its host
star. It was realized quite early that if the stellar tidal dissipation rates inferred from binary star systems were applicable in WASP-18, the system would undergo orbital decay on a timescale of only megayears (\citealt{hellier2009} and references therein). \citet{wilkins2017} recently searched for orbital decay of WASP-18b, concatenating previous observations with new data. Within the context of the ``constant phase lag'' model for tidal interaction \citep{zahn1977}, they reported a limit on the modified stellar tidal quality factor $Q_\star' \geq 1\times10^6$, at 95\% confidence.

\begin{figure*}[t]
	\begin{center}
		\leavevmode
		\includegraphics[width=0.47\textwidth]{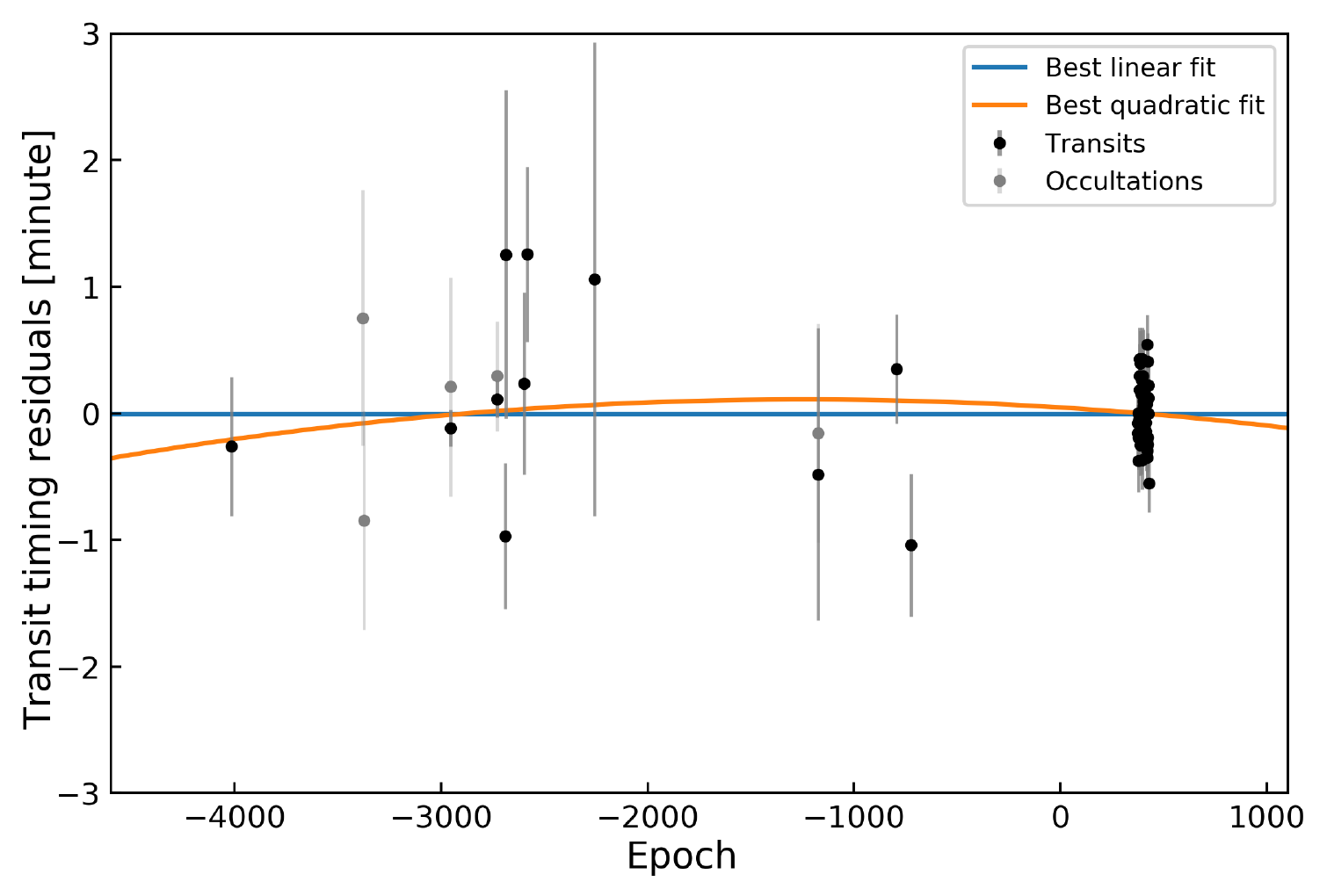}
		\includegraphics[width=0.47\textwidth]{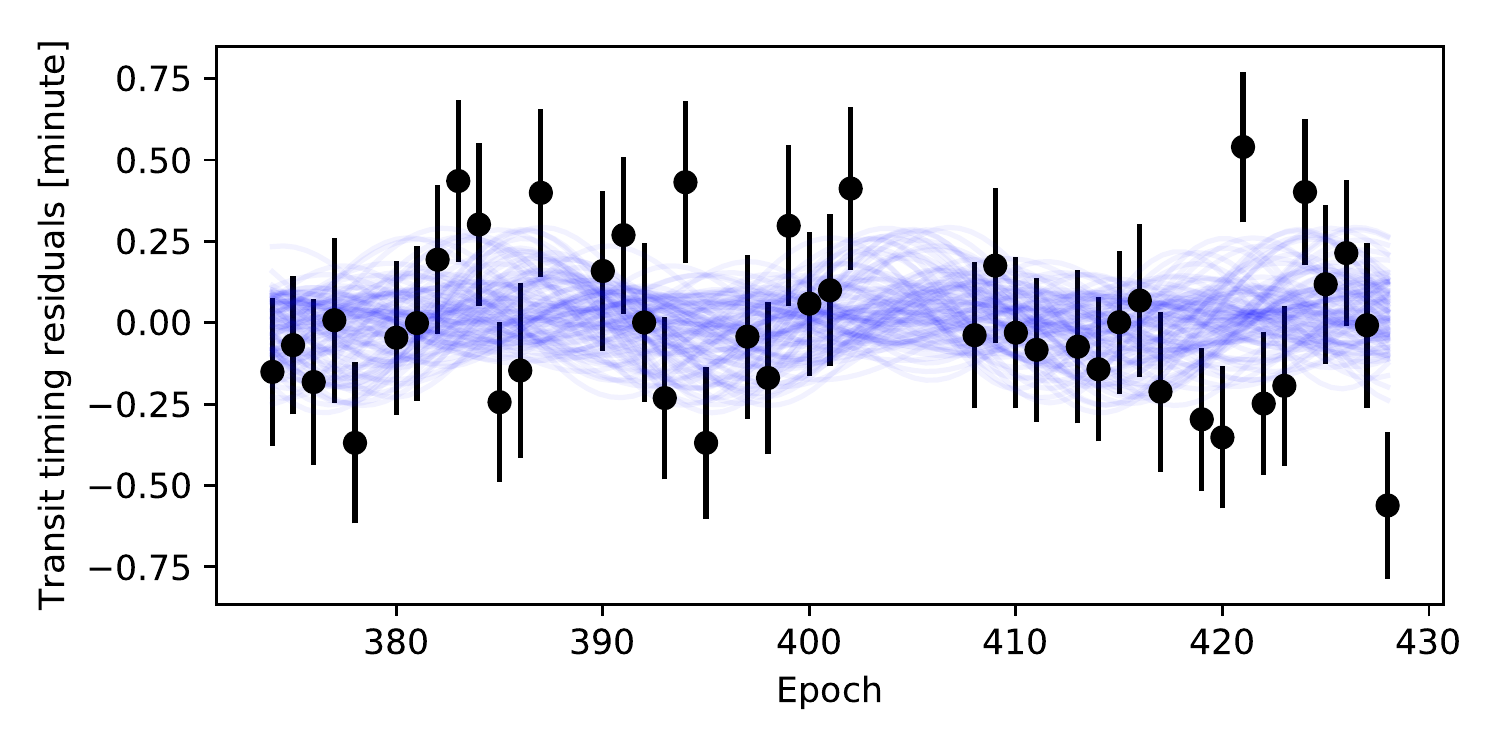}
	\end{center}
	\vspace{-0.7cm}
	\caption{Long-term (left panel) and short-term (right panel) TTV search. Both panels show the residual timing after subtracting the linear ephemeris fitted to the decade-long transit timing data set listed in \tabr{tt}. In the left panel, the best-fit constant-period model (blue) and constant period derivative model (orange) provide comparable fits to the data, but the latter model has one extra free parameter. In the right panel, the blue curves are fair samples drawn from the posterior of a quasi-periodic model used to fit the measured timing residuals. 
    }
	\label{fig:O_minus_C}
\end{figure*}

The \tess\ observations provide new transit times that let us extend WASP-18b
orbital decay search. \tabr{tt} lists the transit and occultation times we used in our long-term timing analysis, where most of the archival times were already compiled by \citet{wilkins2017}. We require that each archival time (i) originate from the peer-reviewed literature, and (ii) be based on observations of a single transit or occultation observed in its entirety. We adopt the methods and equations described by \cite{bouma2019}, who performed a similar study in the context of WASP-4b. To measure the \tess\ transit times, each individual transit is isolated to a window of $\pm 3$ transit durations. Each transit window is then fitted simultaneously for a local linear trend in relative flux, the mid-transit time, and the depth. The remaining transit parameters are fixed to those found in \tabr{modelparams}. Given the abundance of transits, we omitted three \tess\ transits that had significant gaps due to momentum wheel dumps, close to BJDs of 2,458,359.2, 2,458,376.1, and 2,458,390.2.
After deriving the new transit times and uncertainties, we fitted the full timing dataset with two competing models: a linear ephemeris model and a quadratic ephemeris model. \figr{O_minus_C} left panel shows the times and best-fitting models where the linear ephemeris was subtracted. The difference in the Bayesian Information Criterion between the best-fitting linear and quadratic models is $-3.8$, so there is no evidence to prefer the quadratic model over the linear model. The relevant limits from the quadratic model, at 95\% confidence, can be expressed as:
\begin{align}
  -8.78 < \dot{P}  &< 4.98 \ \mathrm{ms}\,\mathrm{year}^{-1}, \label{eq:pdot} \\
  Q_\star' &> 1.73 \times 10^6. \label{eq:q}
\end{align}
As shown in \eqr{pdot} our analysis does not find a long term period derivative. And, as seen in \eqr{q} we do not find the stellar quality factor of $Q\star' \sim 5\times10^5$ suggested by \citet{mcdonald2018}, while our result is a modest improvement relative to the results of \citet{wilkins2017}.

The linear ephemeris we derived by fitting the decade-long transit timings data set is:
\begin{align}
& P = 0.941452419 \pm 0.000000021 \, {\rm day}, \\
& T_0 \, ({\rm BJD_{TDB}}) = 2458002.354726 \pm 0.000023.
\end{align}

\input{WASP-18b_transit_time_table.tex}
%

\subsubsection{Short term TTV}
\label{sec:ttvshort}

The existence of a third body in a star-planet system such as WASP-18 may introduce deviations of the observed transit timings from the expected timings assuming a Keplerian orbit \citep[e.g.,][]{agol2005, holman2005}. Generally known as TTVs, these data features can allow one to probe the system for planetary companions. While these variations can have a wide range of dependencies on the orbital and planetary parameters, the expected generic shape of the TTVs is quasi-sinusoidal.

We analyzed the 44 transit timing inferred from \tess\ Sectors 2 and 3 data, listed in \tabr{tt}, to test whether the observed timings contain any evidence for sinusoidal variations. Towards this purpose, we subtract from observed timings the expected timings based on the linear ephemeris derived in \secr{ttvlong}. Those timing residuals are plotted in \figr{O_minus_C} right panel and we attempt to model those data by sampling from the posterior probability distribution of two TTV models. The first model (the null model) has only a single parameter which is an offset fitted to the timing residuals. The second model has four parameters where in addition to the timing offset it includes the period, phase, and amplitude of the sinusoidal TTV. 

In order to perform the sampling and calculate the associated Bayesian evidences, we use \texttt{dynesty}\footnote{\url{https://github.com/joshspeagle/dynesty}}. The resulting Bayesian evidence of the sinusoidal TTV model relative to that of the null model is -3, indicating that the null hypothesis should be readily preferred. Furthermore, the posterior-mean reduced $\chi^2$ (i.e., root mean square of the residuals per degree of freedom) of the null and alternative models are 1.18 and 1.22, respectively. Hence, we conclude that there is no evidence for TTVs in the \tess\ Sectors 2 and 3 data of WASP-18b.

\section{Summary}
\label{sec:sum}

We have presented here an analysis of the full orbital phase curve of the WASP-18 system measured by \tess\ in Sectors 2 and 3. The per-point residual scatter of the 2-minute data is 520--530~ppm, yielding a binned scatter of 130--160~ppm per 1-hour exposure. This noise level is consistent with the expected noise level of \tess\ data for this $T$=8.83~mag target.

We detect at high significance beaming and ellipsoidal modulations that are consistent with theoretical predictions. We robustly measure a secondary eclipse depth of $341_{-18}^{+17}$ ppm, which when combined with the expected thermal emission in the \tess\ bandpass leads to a null detection of reflected light and a \sig{2} upper limit of 0.048 on the geometric albedo in the \tess\ bandpass. The low optical geometric albedo is consistent with that of other hot Jupiters in similar wavelength range, especially highly-irradiated hot Jupiters \citep{schwartz2015}.

We do not detect a phase shift in the atmospheric phase curve component, with an upper limit of 2.9~deg (\sig{2}), indicating that the phase of maximum light is well-aligned with the phase of secondary eclipse. In addition, we do not detect light from the planet's night-side hemisphere, with an upper limit of 43~ppm (\sig{2}), or 13\% of the day-side brightness. These findings indicate very inefficient distribution of incident energy from the day-side hemisphere to the night-side hemisphere and are consistent with results based on previously published phase curve measurements (both in the near-infrared and in the optical) of similarly highly-irradiated hot Jupiters \citep[e.g.,][]{wong2015,wong2016} and with theoretical expectations \citep{perez2013,komacek2017}.

The clear detection of the WASP-18 phase curve modulations demonstrate that \tess\ data are sensitive to the photometric variations of systems with massive short period planets, a sensitivity that increases when data from several \tess\ sectors is combined. For such objects, the \tess\ phase curve can be used for atmospheric characterization, as shown here, as well as independent mass estimates from the constraints derived from RV analyses.

To complement our study of the phase curve we have also searched for TTVs, both long term using all available measured transit times spanning about a decade,  and also short term within the time span of the \tess\ Sectors 2 and 3 data. In both cases we do not find a statistically significant deviation of the transit timings from a linear ephemeris.

\acknowledgments

We acknowledge the use of \tess\ Alert data, which is currently in a beta test phase. These data are derived from pipelines at the \tess\ Science Office and at the \tess\ Science Processing Operations Center.
Funding for the \tess\ mission is provided by NASA's Science Mission directorate.
This paper includes data collected by the \tess\ mission, which are publicly available from the Mikulski Archive for Space Telescopes (MAST).
Resources supporting this work were provided by the NASA High-End Computing (HEC) Program through the NASA Advanced Supercomputing (NAS) Division at Ames Research Center.
This research has made use of the NASA Exoplanet Archive, which is operated by the California Institute of Technology, under contract with NASA under the Exoplanet Exploration Program.
I.W.~is supported by a Heising-Simons \textit{51 Pegasi b} postdoctoral fellowship.
C.X.H. and M.N.G.~acknowledge support from MIT's Kavli Institute as a Torres postdoctoral fellow.
T.D.~acknowledges support from MIT's Kavli Institute as a Kavli postdoctoral fellow.

{\it Facilities:} 
\facility{\textit{TESS}}

{\it Software:} 
\texttt{astrobase} \citep{bhatti_astrobase_2018},
\texttt{astropy} \citep{astropy2018},
\texttt{BATMAN} \citep{kreidberg_batman_2015},
{\scshape dynesty} (\url{https://github.com/joshspeagle/dynesty}),
\texttt{emcee} \citep{emcee},
\texttt{numpy} \citep{vanderwalt2011}, 
\texttt{matplotlib} \citep{hunter2007}, 
\texttt{pandas} \citep{mckinney-proc-scipy-2010},
\texttt{scipy} \citep{jones_scipy_2001}.




\end{document}

%% file: WASP-18b_transit_time_table.tex
\begin{deluxetable}{ccccc}
    

\tabletypesize{\scriptsize}

\tablecaption{\label{tab:tt} WASP-18b transit \& occultation times}

\tablenum{3}

\tablehead{
  \colhead{$t_{\rm mid}$ [BJD$_\mathrm{TDB}$]} &
  \colhead{$\sigma_{t_{\rm mid}}$ [days]} &
  \colhead{Epoch} & 
  \colhead{Reference}
}

\startdata
 2454221.48163 &      0.00038 &   -4017 &    \citet{hellier2009} \\
 2454820.71680 &      0.00070 &   -3380.5 &     \citet{nymeyer2011} \\ 
 2454824.48150 &      0.00060 &   -3376.5 &     \citet{nymeyer2011} \\
 2455220.83370 &      0.00060 &   -2955.5 &     \citet{maxted2013} \\
 2455221.30420 &      0.00010 &   -2955 &     \citet{maxted2013} \\
 2455431.71910 &      0.00030 &   -2731.5 &     \citet{maxted2013} \\
 2455432.18970 &      0.00010 &   -2731 &     \citet{maxted2013} \\
 2455470.78850 &      0.00040 &   -2690 &     \citet{maxted2013} \\
 2455473.61440 &      0.00090 &   -2687 &     \citet{maxted2013} \\
 2455554.57860 &      0.00050 &   -2601 &     \citet{maxted2013} \\
 2455570.58400 &      0.00048 &   -2584 &     \citet{maxted2013} \\
 2455876.55590 &      0.00130 &   -2259 &     \citet{maxted2013} \\
 2456895.67730 &      0.00060 &   -1176.5 &     \citet{wilkins2017} \\
 2456896.14780 &      0.00080 &   -1176 &     \citet{wilkins2017} \\
 2457255.78320 &      0.00030 &    -794 &     \citet{wilkins2017} \\
 2457319.80100 &      0.00039 &    -726 &     \citet{wilkins2017} \\
 2458354.45782 &      0.00016 &     374 &                       This work \\
 2458355.39933 &      0.00015 &     375 &                       This work \\
 2458356.34070 &      0.00018 &     376 &                       This work \\
 2458357.28229 &      0.00018 &     377 &                       This work \\
 2458358.22348 &      0.00017 &     378 &                       This work \\
 2458360.10661 &      0.00016 &     380 &                       This work \\
 2458361.04809 &      0.00017 &     381 &                       This work \\
 2458361.98968 &      0.00016 &     382 &                       This work \\
 2458362.93130 &      0.00017 &     383 &                       This work \\
 2458363.87266 &      0.00017 &     384 &                       This work \\
 2458364.81373 &      0.00017 &     385 &                       This work \\
 2458365.75526 &      0.00019 &     386 &                       This work \\
 2458366.69709 &      0.00018 &     387 &                       This work \\
 2458369.52128 &      0.00017 &     390 &                       This work \\
 2458370.46281 &      0.00017 &     391 &                       This work \\
 2458371.40407 &      0.00017 &     392 &                       This work \\
 2458372.34536 &      0.00017 &     393 &                       This work \\
 2458373.28728 &      0.00017 &     394 &                       This work \\
 2458374.22817 &      0.00016 &     395 &                       This work \\
 2458376.11131 &      0.00017 &     397 &                       This work \\
 2458377.05267 &      0.00016 &     398 &                       This work \\
 2458377.99445 &      0.00017 &     399 &                       This work \\
 2458378.93573 &      0.00015 &     400 &                       This work \\
 2458379.87722 &      0.00016 &     401 &                       This work \\
 2458380.81889 &      0.00017 &     402 &                       This work \\
 2458386.46729 &      0.00016 &     408 &                       This work \\
 2458387.40889 &      0.00017 &     409 &                       This work \\
 2458388.35020 &      0.00016 &     410 &                       This work \\
 2458389.29161 &      0.00015 &     411 &                       This work \\
 2458391.17453 &      0.00016 &     413 &                       This work \\
 2458392.11593 &      0.00015 &     414 &                       This work \\
 2458393.05748 &      0.00015 &     415 &                       This work \\
 2458393.99898 &      0.00016 &     416 &                       This work \\
 2458394.94024 &      0.00017 &     417 &                       This work \\
 2458396.82309 &      0.00015 &     419 &                       This work \\
 2458397.76450 &      0.00015 &     420 &                       This work \\
 2458398.70657 &      0.00016 &     421 &                       This work \\
 2458399.64748 &      0.00015 &     422 &                       This work \\
 2458400.58897 &      0.00017 &     423 &                       This work \\
 2458401.53083 &      0.00016 &     424 &                       This work \\
 2458402.47209 &      0.00017 &     425 &                       This work \\
 2458403.41361 &      0.00016 &     426 &                       This work \\
 2458404.35491 &      0.00018 &     427 &                       This work \\
 2458405.29598 &      0.00016 &     428 &                       This work \\
\enddata


\tablecomments{
    $t_{\rm mid}$ is the midtime of the transit or occultation;
    $\sigma_{t_{\rm mid}}$ is its $1\sigma$ uncertainty.
    The ``Reference'' column indicates the work describing the
    original observations.
    The literature times are all collected from the homogeneous
    \citet{wilkins2017} analysis.
    Occultation times have been corrected for the light travel time
    across the system (cf. \citealt{maxted2013}).
}

\end{deluxetable}